\documentclass[]{agujournal}
\usepackage{aas_macros}
\usepackage[caption=false]{subfig}
\draftfalse

\newcommand{\vect}[1]{\mathbf{#1}}
\newcommand{\HI}{H\,{\sc i}}

\journalname{Radio Science}

\begin{document}

% Example: \title{This is a test title}

\title{\HI\ 21~cm Cosmology and the Bi-spectrum: Closure Diagnostics in Massively Redundant Interferometric Arrays}

% Example: \authors{A. B. Author\affil{1}\thanks{Current address, Antartica}}
% Author\affil{3,4}\thanks{Also funded by Monsanto.}}

\authors{
% primary authors
C.L. Carilli\affil{1,2}, 
Bojan Nikolic\affil{2}, 
Nithyanandan Thyagarajan\affil{1,3}\thanks{Nithyanandan Thyagarajan is a Jansky Fellow of the National Radio Astronomy Observatory.}, 
K. Gale-Sides\affil{2}, 
% Builders list
Zara  Abdurashidova\affil{5},
James E. Aguirre\affil{4}, 
Paul Alexander\affil{2}, 
Zaki S. Ali\affil{5}, 
Yanga  Balfour\affil{13},
Adam P. Beardsley\affil{3},
Gianni Bernardi\affil{11,12}, 
Judd D. Bowman\affil{3}, 
Richard F. Bradley\affil{10},
Jacob  Burba\affil{6},
Carina Cheng\affil{5}, 
David R. DeBoer\affil{5}, 
Matt  Dexter\affil{5},
Eloy de~Lera~Acedo\affil{2}, 
Joshua S. Dillon\affil{5}, 
Aaron Ewall-Wice\affil{9}, 
Gcobisa Fadana\affil{13}, 
Nicolas Fagnoni\affil{2}, 
Randall Fritz\affil{13}, 
Steve R. Furlanetto\affil{7},
Abhik Ghosh\affil{13},
Brian Glendenning\affil{1}, 
Bradley Greig\affil{12}, 
Jasper Grobbelaar\affil{13}, 
Ziyaad  Halday\affil{13},
Bryna J. Hazelton\affil{14,15}, 
Jacqueline N. Hewitt\affil{9}, 
Jack Hickish\affil{5}, 
Daniel C. Jacobs\affil{3},
Austin Julius\affil{13}, 
MacCalvin Kariseb\affil{13}, 
Saul A. Kohn\affil{4},
Mathew Kolopanis\affil{3},
Telalo Lekalake\affil{13}, 
Adrian Liu\affil{5}\thanks{A. Liu is a Hubble Fellow at UCB}, 
Anita Loots\affil{13}, 
David MacMahon\affil{5}, 
Lourence Malan\affil{13}, 
Cresshim Malgas\affil{13}, 
Matthys Maree\affil{13}, 
Zachary Martinot\affil{4},
Eunice Matsetela\affil{13}, 
Andrei Mesinger\affil{12},
Mathakane  Molewa\affil{13},
Miguel F. Morales\affil{14}, 
Abraham R. Neben\affil{9}, 
Aaron R. Parsons\affil{5}, 
Nipanjana Patra\affil{5},
Samantha Pieterse\affil{13}, 
Paul La Plante\affil{4}, 
Jonathan C. Pober\affil{6},
Nima Razavi-Ghods\affil{2},
Jon Ringuette\affil{14},
James Robnett\affil{1},
Kathryn Rosie\affil{13}, 
Raddwine Sell\affil{13}, 
Peter  Sims\affil{6},
Craig Smith\affil{13}, 
Angelo Syce\affil{13},
Peter K.~G. Williams\affil{8},
Haoxuan Zheng\affil{9}
}

\affiliation{1}{National Radio Astronomy Observatory, P. O. Box 0, Socorro, NM 87801, USA}
\affiliation{2}{Astrophysics Group, Cavendish Laboratory, JJ Thomson Avenue, Cambridge CB3 0HE, UK}
\affiliation{3}{Arizona State University, School of Earth and Space Exploration, Tempe, AZ 85287, USA}
\affiliation{4}{Department of Physics and Astronomy, University of Pennsylvania, Philadelphia, PA} 
\affiliation{5}{Department of Astronomy, University of California, Berkeley, CA, USA}
\affiliation{6}{Physics Department, Brown University, Providence, RI}
\affiliation{7}{Department of Physics and Astronomy, University of California, Los Angeles, CA}
\affiliation{8}{Harvard-Smithsonian Center for Astrophysics, Cambridge, MA} 
\affiliation{9}{Department of Physics, Massachusetts Institute of Technology, Cambridge, MA}
\affiliation{10}{National Radio Astronomy Observatory, Charlottesville, VA} 
\affiliation{11}{Department of Physics and Electronics, Rhodes University, Grahamstown, South Africa} 
\affiliation{12}{Scuola Normale Superiore, Pisa, Italy} 
\affiliation{13}{SKA-SA, Cape Town, South Africa}
\affiliation{14}{Department of Physics, University of Washington, Seattle, WA} 
\affiliation{15}{eScience Institute, University of Washington, Seattle, WA}

% Example: \correspondingauthor{First and Last Name}{email@address.edu}

\correspondingauthor{Chris Carilli}{ccarilli@nrao.edu}

%% Keypoints, final entry on title page.

\begin{keypoints}
\item New massively redundant low frequency arrays allow for a novel investigation of closure relations in radio interferometry.
\item We use closure phase spectra for redundant triads to estimate departures from redundancy for redundant baseline visibilities. 
\item We investigate the temporal behaviour of closure spectra, and show that time averaging should be limited to about 1min to 2min, due to transit of the sky through the primary beam of the telescope.   
\item We show that closure phase spectra can also be used to identify bad antennas in the array, independent of calibration. 
\item We develop the tools and framework in which closure phase spectra may be used in the search for the H~{\sc i} 21~cm signal from cosmic reionization.
\end{keypoints}

%% ------------------------------------------------------------------------ %%
%
%  ABSTRACT
%
%% \begin{abstract} starts the second page 

\begin{abstract}

New massively redundant low frequency arrays allow for a novel
investigation of closure relations in interferometry. We employ
commissioning data from the Hydrogen Epoch of
Reionization Array to investigate closure quantities in this densely
packed grid array of 14m antennas operating at 100 MHz to 200 MHz. We  
investigate techniques that utilize
closure phase spectra for redundant triads to estimate departures from redundancy for redundant baseline visibilities. We find a median 
absolute deviation from redundancy in closure phase across the observed frequency range of about $4.5^\circ$.  This value translates 
into a non-redundancy per visibility phase of about 
$2.6^\circ$, using proto-type electronics. 
The median absolute deviations from redundancy decrease with longer
baselines. We show that 
closure phase spectra can be used to identify ill-behaved 
antennas in the array, independent of calibration.  We investigate the 
temporal behaviour of closure spectra. The Allan variance increases
after a one minute stride time,  due to passage of the sky through the 
primary beam of the transit telescope. However, the closure spectra 
repeat to well within the noise per measurement at corresponding local 
sidereal times (LST) from day to day. In future papers in this series we will
develop the technique of using closure phase spectra in the search for the 
\HI\ 21~cm signal from cosmic reionization.
  
\end{abstract}

%% ------------------------------------------------------------------------ %%
%
%  TEXT
%
%% ------------------------------------------------------------------------ %%

\section{Introduction} 

Closure quantities have long been a tool in interferometric
imaging \citep{jen58, rog74}.
As is well known, calculation of closure quantities from interferometric 
visibilities is independent of phase and amplitude terms introduced by 
antenna optics and electronics (e.g., standing waves), as well as to 
electrical path length and attenuation contributions due to the 
ionosphere  and troposphere above the antenna. Hence, closure quantities 
represent a robust measurement of the sky signal, independent of 
antenna-based calibration and calibration errors \citep{joh17, roe17}.

This is the first in a series of papers considering closure
measurements, and their application to \HI\ 21~cm cosmology. Low
frequency radio astronomy is undergoing a renaissance due to the
intensive search for the 21~cm line signal from neutral Hydrogen during
the epoch of cosmic reionization, and into the preceding dark ages
\citep{fan06,mor10}.  
These low frequency measurements are complicated by the strong
continuum sky signal, coupled to any chromatic response of the telescope.
Examination of closure spectra may provide a robust method to detect
the \HI\ signal, independent of antenna-based calibration and the
ionosphere \citep{car16a, tcn18}.
Study of the closure spectra also provides powerful, calibration-independent, 
diagnostics of the array performance \citep{car17}.

We develop the formalism and tools for making the
closure measurements in the context of the massively redundant
interferometric transit array, the Hydrogen Epoch of Reionization
Array \citep[HERA;][]{deb17,ewa16,neb16,patra17,thy16}.
HERA provides us a unique laboratory to explore closure as an array 
diagnostic, including measuring the level of redundancy for the baselines, 
searching for ill-behaved antennas, and measuring the effect on 
coherent averaging of time variation as the sky transits the array.  In 
the second paper, we will present the cosmological 
formalism for searching for the \HI\ 21~cm signal using closure 
measurements \citep{tcn18}. 
In a third paper, we will apply the technique to HERA 
observations taken over the 2017-2018 winter. 
We emphasize at the
start that there are non-closing effects, possibly due to polarization leakage,
cross coupling of antennas, and non-identical far sidelobes of the telescope
beams that can lead to departures from the closure relationship. We consider
these in more detail below.

\section{\HI\ 21~cm cosmology: context and challenges}

Detecting the \HI\ 21~cm line signal from the neutral intergalactic
medium during cosmic reionization 
is one of the paramount goals of modern astrophysics 
\citep{sun72,sco90,mad97,toz00,ili02,fan06,bar07,mor10}.
These epochs correspond to the early formation of galaxies, 
within a few hundred million years of the Big Bang. Current indirect 
measurements suggest cosmic reionization occurred at $z \sim$~6--10 
\citep{ban17,rob15,gre17}. Hence, the expected \HI\ 21~cm signal from the 
epoch of reionization (EoR) will be observed at frequencies from 100~MHz to 
200~MHz. Most recently, there is a possible detection of the all-sky
HI 21cm signal from the IGM at $z  \sim 18$, during the so-called 'cosmic
dawn', when the very first generation of stars form \citep{bow18, bar18}.

Numerous low frequency interferometers and total power experiments
have been built with the thermal sensitivity adequate to detect
the \HI\ 21~cm signal from the reionization epoch \citep{thy13,bea13, mon17}.  However, the task is 
complicated by the much stronger foreground continuum emission: the 
required dynamic range is about 50dB in the mean brightness 
\citep{mon17}, and 40dB for spatial fluctuations on the scales relevant to current 
interferometric observations \citep{sim16}.

A crucial distinguishing property of the foregrounds is that they 
are dominated by synchrotron radiation. Synchrotron radiation 
is a non-thermal process, which is well determined in
the laboratory and in space to display power-law spectra  
over a very broad frequency range. Such power-law spectra 
are featureless over scales of hundreds of MHz
\citep{pac70}. This spectral smoothness is in stark contrast to the \HI\ 21~cm line 
emission, which fluctuates rapidly in frequency down to kHz scales
\citep{car02}. A naive solution to remove the continuum foregrounds would be fitting a 
smoothly varying function (such as a polynomial) in frequency to either 
the visibilities or the spectral image cubes. However, \citet{sim16} 
show that, without
{\it a priori} knowledge of the covariance between the foregrounds and 
the 21~cm signal in the data, blind subtraction of a foreground
model from the data prior to estimation of the quantity of interest
will produce biased estimates of said quantity. As such, joint
estimation of the foregrounds and 21~cm signal is essential for
obtaining statistically robust estimates of the 21~cm signal.

Perhaps more importantly, \citet{dat09,dat10}
show that the chromatic response of an interferometer for very wide-field 
imaging imprints a spectral signature on the visibility data which is
impossible to remove using continuum subtraction techniques via smooth
curve-fitting to either the visibility spectra,
or position-by-position smooth curve-fitting to
a spectral image cube \citep{rup99}.

The continuum can still be removed properly, in theory, through a 
frequency-dependent subtraction from the visibilities of an accurate 
continuum model.
The challenge, as \citet{dat10}
show, is that such a subtraction requires remarkably accurate complex gain 
calibration ($\sim 0.1\%$), as a function of frequency \citep{dat09,dat10}.
These techniques, as exemplified for example by LOFAR analysis of the 
reionization signal, are known as `foreground removal techniques' 
\citep{patil17}. 
Progress has been made on foreground removal techniques, using
the spectral and spatial characteristics, in particular
by \citep{tro16, mer17}, however, a detection of the statistical fluctuations
of the HI 21cm signal has not yet been made.

These facts have led to consideration of alternate methods for detecting 
the \HI\ 21~cm signal, such as `foreground avoidance' in delay spectra 
\citep{mor12,par12}.
The delay spectrum method involves separating \HI\ line from continuum 
emission in the three-dimensional power spectral space defined by the line 
of sight (frequency, or equivalently, redshift) distances and the sky-plane 
(angular) distances.  In this space, the maximum wavenumber (or spectral 
frequency) for smooth-spectrum continuum emission due to the chromatic 
response of a given interferometric baseline is set by the maximum delay of 
the baseline corresponding to sources at the horizon. Hence, the HI 21cm line 
signal in the line-of-sight direction appears at all wavenumbers, including large 
wavenumbers (small scales), whereas the continuum emission is limited to a `wedge' of 
low wavenumbers determined by the horizon limit, as demonstrated by \citet{dat10} 
and analyzed by \citet{mor12,par12,tro12,ved12,pob13,thy13,liu14a}.
 
The avoidance method still requires tight control of the spectral response 
of other parts of the array, such as the antennas and data transmission 
system, and/or very accurate calibration of the spectral response (bandpass) 
with time, wide-field effects, spectral structures in the antenna's angular 
power pattern, and reflections, to avoid coupling the continuum signal to the 
line, and thereby causing leakage of the continuum signal into the 
EoR window \citep[e.g.,][]{thy15a,thy15b,thy16,pob16}.

In our broader study, we consider an approach for discovering the \HI\ 21~cm signal using 
the closure phases of the interferometer. A closure phase is a quantity derived 
from a simple product of the three
visibility pairs from three antennas, more generally known as the
`bispectrum'\footnote{The bispectrum is a well known statistical tool
to study e.g., cosmological anisotropies \citep{son15}.
However, such cosmological analyses only consider the scalar
amplitudes of the bispectrum, not the vector phases \citep{has17}.} 
\citep{jen58,kul89,mon07}. It was recognized early in the field of both optical
and radio interferometry that closure quantities are independent of
antenna-based contributions to the measured phases and amplitudes,
such as standing waves arising in antenna structures or cables. This
independence also applies to the contribution to electrical path length
by atmospheric fluctuations above each antenna (principally due to
the troposphere in the case of optical or high frequency radio
observations, and the ionosphere at low radio frequencies), 
although such fluctuations are small on the very short baselines of HERA.

Hence, to the degree that the array response is separable into
antenna-based terms, closure phases are independent of calibration and
calibration errors, as well as independent of the ionosphere,
i.e., closure quantities are a robust `observable' of the true sky
signal. This fact was used in early radio interferometry. In
particular, closure analysis was an enabling technique in early Very Long Baseline Interferometry (VLBI) experiments, where
maintaining phase coherence was problematic \citep{pea84,wil78,rog74},
and remains crucial in current submm VLBI analyses \citep{joh17,roe17}.
Closure quantities are still used extensively
in optical interferometric image reconstruction \citep{thi10},
as well as being a primary diagnostic
for antenna-based calibration errors in phase-connected radio
interferometers \citep{per99}.

\section{Closure Phase Review}

We briefly review the definition of closure phase \citep[see][for more
detail]{cor99,tho17}.

\subsection{Mathematical foundation}

A radio-interferometer measures the time-averaged cross correlation of
the electric field voltages from pairs of antennas. Setting aside the
polarization of the electric field for the moment, the
quasi-monochromatic response of the interferometer can be derived from
the van Cittert-Zernike theorem as given by \cite{tho17}:
\begin{equation}
  \mathcal{V}_{ab}=\int d\Omega F_a(\vect{\sigma}) F^{*}_b(\vect{\sigma}) I(\vect{\sigma}) e^{-j2\pi \vect{d}_{ab}\cdot\vect{s}}
  \label{eq:visdef}
\end{equation}
where $\mathcal{V}_{ab}$ is the response (`visibility') measured
between antennas number $a$ and $b$, $d\Omega$ is an infinitesimal
element of solid angle, $F_a(\vect{s})$ is the antenna voltage
response pattern of antenna $a$ in direction of the unit vector $\vect{s}$,
$I(\vect{s})$ is the intensity of radiation from direction
$\vect{s}$, and $\vect{d}_{ab}$ is the vector of between positions
of antenna $a$ and $b$ measured in wavelengths of the radiation being
received. In practice, any measurement will be an integral over a
finite frequency band and sum over polarization states of the
incident field.

The `bi-spectrum' or `triple product' for an interferometric
measurement for antennas $a$, $b$ and $c$ is defined as:
\begin{equation}
  \mathcal{C}_{abc}=\mathcal{V}_{ab} \mathcal{V}_{bc} \mathcal{V}_{ca}.
\end{equation}
There are two regimes in which this quantity has particularly simple
properties and from which stems its practical use.

The first regime of interest is when there is a single point source in
direction $\vect{s}$ which dominates the received incoming radiation;
in this case the integral Eq~(\ref{eq:visdef}) reduces to:
\begin{equation}
  \mathcal{V}_{ab}=  F_a(\vect{s}) F^{*}_b(\vect{s}) I_\vect{s} e^{-j2\pi \vect{d}_{ab}\cdot\vect{s}}  
\end{equation}
and the triple product is:
\begin{eqnarray}
  \mathcal{C}_{abc} &=& \left| F_a(\vect{s})  \right| ^2  \left| F_b(\vect{s})  \right| ^2  \left| F_c(\vect{s})  \right| ^2
                        e^{-j2\pi (\vect{d}_{ab} + \vect{d}_{bc} +\vect{d}_{ca})\cdot\vect{s}}  \\
  &=&  \left| F_a(\vect{s})  \right| ^2  \left| F_b(\vect{s})  \right| ^2  \left| F_c(\vect{s})  \right| ^2 \label{eq:closurephasezero}
\end{eqnarray}
where Eq~(\ref{eq:closurephasezero}) follows because $\vect{d}_{ab}$,
$\vect{d}_{bc}$ and $\vect{d}_{ca}$ form a closed triangle and
therefore add to null. The phase of $\mathcal{C}_{abc}$ (which is
normally referred to as the `closure phase') is evidently zero in this
case entirely independently of the responses of the individual
antennas.

The second regime of interest is when the responses of each antenna
differ by only a (complex) constant factor that is independent of
direction, i.e.,
$F_{a}(\vect{s}) = F'_{a}\,F (\vect{s})$. The
triple product is then
\begin{eqnarray}
  \mathcal{C}_{abc} &=& \left| F'_a \right| ^2  \left| F'_b \right| ^2  \left| F'_c \right| ^2
                        \int d\Omega \left| F(\vect{s}) \right|^2 I(\vect{s}) e^{-j2\pi \vect{d}_{ab}\cdot\vect{s}} \times \nonumber\\
                   &&\times   \int d\Omega \left| F(\vect{s}) \right|^2 I(\vect{s}) e^{-j2\pi \vect{d}_{bc}\cdot\vect{s}} \times
                        \int d\Omega \left| F(\vect{s}) \right|^2 I(\vect{s}) e^{-j2\pi \vect{d}_{ca}\cdot\vect{s}} \label{eq:closurephasegen}
\end{eqnarray}
It is again evident that the closure phase, i.e, the phase of the
triple product, is independent of the individual antenna responses
$F'_a$, $F'_b$ and $F'_c$. In this case however the phase will not be
zero in general and will depend on the sky brightness distribution and
the angular dependence of the antenna response pattern.

We emphasize that there are non-closing effects that impact the assumptions
made above. We discuss these in depth in \S\ref{sec:nonred}.
There is an analogous quantity, the `closure amplitude' \citep{cor99,joh17},
based on the combination of four visibility measurements for which
\emph{amplitude} is independent of the antenna-based variation in
antenna response. We discuss this closure amplitude briefly in
\S\ref{sec:clamp}.

\subsection{Use of closure phase}

Closure phase is used both as a diagnostic and in inference of the sky
brightness distribution. Its use as a diagnostic depends on the property shown in
Eq~(\ref{eq:closurephasezero}), i.e., when a single point source
dominates a perfect interferometer will measure zero closure phase
regardless of the individual antenna responses. The antenna response
includes the many mechanical, electrical and most atmospheric effects
that affect the response of the antenna to celestial radiation and are
either not known or are time variable.

Most general purpose radio telescopes observing at cm or shorter
wavelengths are made of steerable high-directivity antennas, and the
sky at these wavelengths has pronounced point-like sources, it is
usually possible to make an observation that satisfies the condition
of a single dominant source. The closure phase then provides a direct
diagnostic of the interferometer without need for calibration of the
unknown or time variable effects, and indeed without the need for any
sophisticated post processing of the measured visibilities.

Use of closure phase in constraining models of the sky depends on the
property of Eq~(\ref{eq:closurephasegen}), i.e., that the phase is
independent of any unknown or time-varying properties of the antennas
that are independent of the angle of arrival of the radiation. The
measured closure phase is then a property of the sky and the known
angular dependence of the antenna reception pattern only, and can be
used to constrain models of the sky without the need for calibration. 
This use of closure phase was recognized early in the field of
astronomical interferometry \citep{jen58} and is still often used in
situations where instrumental phase stability and determination of
antenna-based calibration terms may be difficult, such as in certain
historical VLBI applications, and in interferometry at optical
wavelengths.

We explore the use of closure phase as a diagnostic when
the condition of a single dominant point source \emph{is not} met. 
In this case we do not know \emph{a-priori} the expected closure phase. 
We can however expect that redundant triads should measure the \emph{same}
closure phase. More precisely
$ \mathcal{C}_{abc} = \mathcal{C}_{def} $ if $\vect{d}_{ab} = \vect{d}_{de}$,
$\vect{d}_{bc} = \vect{d}_{ef}$ and $\vect{d}_{ca} = \vect{d}_{fd}$. 
In the new generation of highly-redundant interferometric arrays, there are many such 
redundant triads.

A real interferometer will introduce both thermal noise and complex
gain terms, $G_i$ (amplitude and phase terms due to the instrument
response; note that this response includes path length differences due to
the ionosphere above each antenna \citep{mev16}, that will alter the sky visibility,
resulting in a measured quantity, $\mathcal{V}_{i,j}^\textrm{m} (u,v)$:

\begin{eqnarray}\label{eqn:obsvis}
\mathcal{V}_{i,j}^\textrm{m} &= G_i G_j^\ast \mathcal{V}_{i,j}^\textrm{s} = a_i e^{i\theta_i} a_j e^{-i\theta_j} A^s_{i,j} e^{i\phi_{i,j}^s} \quad +\quad N_{i,j}
\end{eqnarray}

\noindent where $A^s_{i,j}$ is the true sky visibility amplitude,
$\phi_{i,j}^s$ is the sky visibility phase, and 
$\theta_i$ is the phase introduced to the visibility
by the antenna electronics, optics, or ionosphere, 
$N_{i,j}$ is the noise added to 
the visibility, $\mathcal{V}_{i,j}^\textrm{s}$ is the 
effective sky visibility\footnote{The true sky visibility set by the 
sky intensity distribution multiplied by the
primary beam power pattern
$\equiv A^s_{i,j} e^{i\phi_{i,j}^s}$}, and $a_i$ is the gain amplitude of the antenna plus 
electronics. This assumes that the complex gain on a given visibility 
is separable into antenna-based terms.

From this, we can see that the resulting measured visibility phase is
the sum of exponents:

\begin{eqnarray}
\phi_{i,j}^\textrm{m} &= \phi_{i,j}^\textrm{s} + (\theta_i - \theta_j) + \phi_{i,j}^\textrm{n}
\end{eqnarray}
where, $\phi_{i,j}^\textrm{n}$ is the noise in the measured visibility phase.

Again, the  `bi-spectrum' or `triple product' for an interferometric measurement
is defined as:

\begin{eqnarray}
C_{i,j,k}^\textrm{m} &= \mathcal{V}_{i,j}^\textrm{m} \mathcal{V}_{j,k}^\textrm{m} \mathcal{V}_{k,i}^\textrm{m}
\end{eqnarray}

\noindent It is easy to see from the equations above that the phase of
this complex measurement, or closure phase, is, again, the sum of exponents,
ie. the sum of the three measured visibility phases:

\begin{eqnarray}
\phi_{i,j,k}^\textrm{m} &= \phi_{i,j}^\textrm{s} + (\theta_i - \theta_j) +
\phi_{j,k}^\textrm{s} + (\theta_j - \theta_k) + \phi_{k,i}^\textrm{s} + (\theta_k -
\theta_i) + \phi_{i,j,k}^\textrm{n}
\end{eqnarray}
where, $\phi_{i,j,k}^\textrm{n}$ is the noise in the measured closure phase.

\noindent The antenna based phase terms then cancel in such a
sum, leading to:

\begin{eqnarray}
\phi_{i,j,k}^\textrm{m} &= \phi_{i,j,k}^\textrm{s} + \phi_{i,j,k}^\textrm{n}.
\end{eqnarray}

\begin{figure}[h]
\centering 
\includegraphics[scale=0.6]{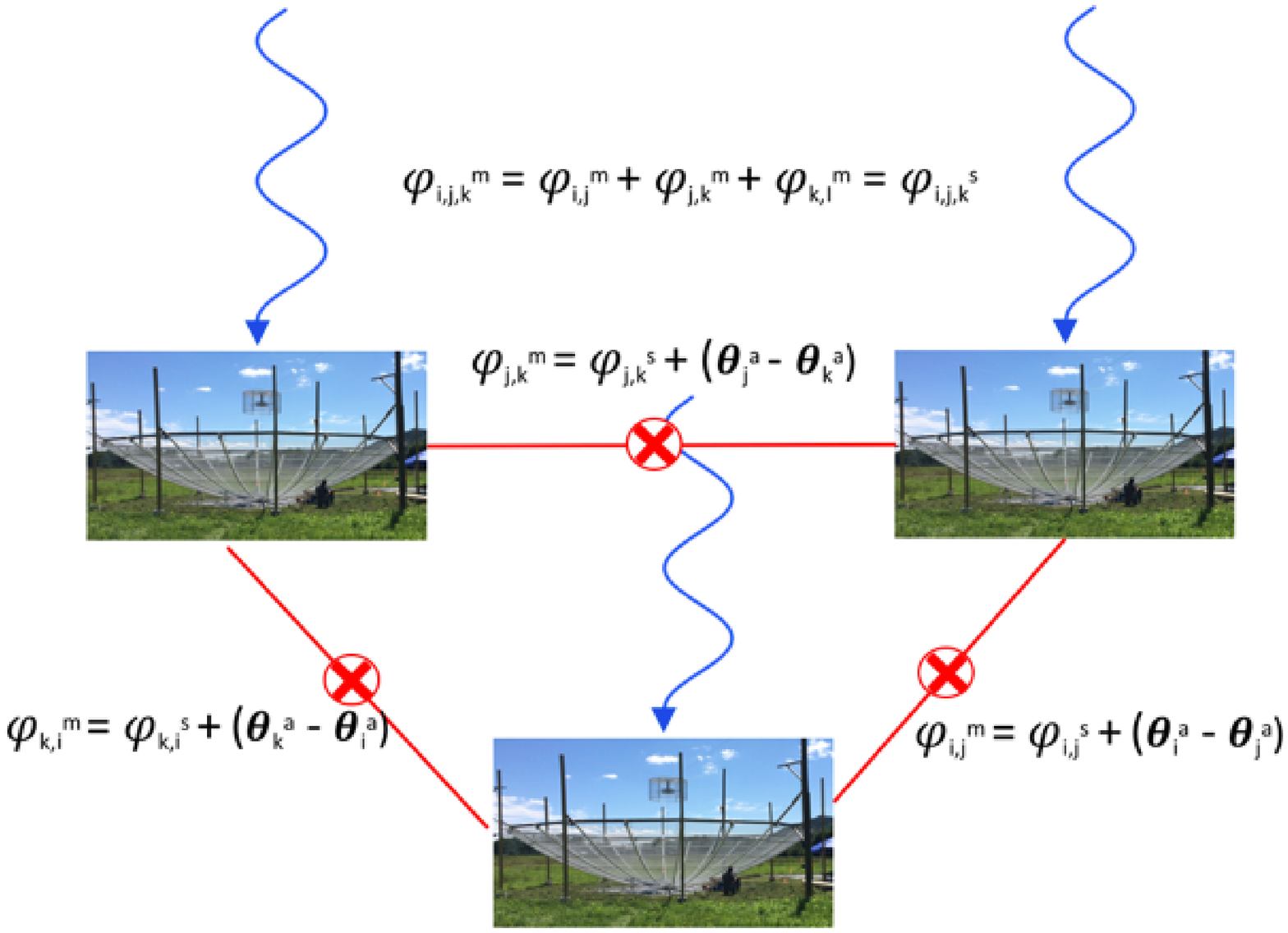}
\caption{Schematic of the closure phase calculation for a three-element interferometer with elements $i,j,k$. $\phi^m$ is the frequency-dependent measured phase of a visibility (cross-correlation of voltages), between two antennas. $\phi^s$ is the true sky value of the phase (i.e., uncorrupted by the response of the system). $\theta^a$ is the electronic phase term introduced by each antenna, including the contribution to the electronic path-length by the atmosphere above each antenna. Assuming that the electronic response of the system can be factorized into antenna-based terms, $\theta^a$, then, in the sum of the measured interferometric phases around any triad of antennas, these antenna-based phase terms cancel, leaving a `true' sky measurement, modulo a thermal noise term that is not shown explicitly.}
\label{fig:schematic}
\end{figure}

\noindent The closure phase is schematically illustrated in Figure~\ref{fig:schematic}.
The implication is that the measured closure phase is {\sl
independent of antenna-based calibration terms of the form
given in the equations above}, and represents a
direct measurement of the true closure phase due to structure on the
sky, modulo the system thermal noise.

The power of the closure phase spectral analysis is that it
is robust to multiplicative contributions to the sky signal which can
be separable into antenna gains, as per the historic standard
interferometric sky calibration process \citep{per99}, 
since these cancel in the closure triad calculation. The closure quantity is a true measure of the sky, independent
of antenna-base gains.  For comparison, redundant visibility spectra are not robust to such antenna-based contributions, and hence are not a true
measure of the sky until after calibration is applied. As an
example, if there were a strong, and varying, standing wave in the
RF/IF bandpass response in one antenna, this will adversely affect the
visibilities, and preclude a sky-based measurement without
first accurate calibration of the standing wave, but this
standing wave would be unseen in the closure analysis. The closure
spectrum remains a true sky measurement. The most recent application
of closure measurements as a robust sky measurement independent of
antenna-base calibration involves interferometric measurements of
the supermassive Black Hole at the Galactic center. These include
millimeter and submm VLBI measurements, in which
time variations of the phase contributions of the 
antenna electronics, and/or the troposphere,
preclude accurate antenna-based calibration \citep{roe17, joh17}.

Of course, this conclusion relies on the assumption that the phase induced by the system is factorizable into antenna-based terms, i.e., that the correlator 
or other aspects of the system do not introduce phase terms that depend on the 
particular cross-correlation for a visibility.  Such non-closing terms are 
known as `closure errors', and remain an important diagnostic of the quality 
of antenna-based calibration in interferometers \citep{fom99}, 
and the ultimate limitation to high dynamic range imaging using 
hybrid-mapping \citep{pea84}.

\section{The HERA Array and Data Processing}

\subsection{The Array}\label{sec:array}

HERA is a close-packed grid of 14~m-diameter parabolic transit
reflectors currently observing in the band 100--200~MHz in the
Karoo Radio Quiet Preserve in South Africa \citep{deb17} at a latitude
of $-30.73^\circ$. The array design is optimized for both the redundant 
spacing calibration technique, and for the delay spectrum technique for 
detection of the H~{\sc i} 21~cm signal from cosmic reionization 
\citep{par12,deb17}. The primary beam FWHM of the dish is about $8^\circ$ at 
150MHz.

The layout of HERA is a split-core hexagonal grid \citep{dil16}, with 
the smallest spacing between two antennas of 14.6~m.  The massively redundant 
HERA configuration allows for a unique application of closure analysis, both 
in terms of array diagnostics and tests of redundancy, and potentially in the
pursuit of H~{\sc i} 21~cm signal detection using closure spectral analysis. 
We establish the closure calculation formalism and consider 
array diagnostics. The cosmological application will be presented in the two 
subsequent papers in the series.

\begin{figure}
\centering 
\includegraphics[scale=0.5]{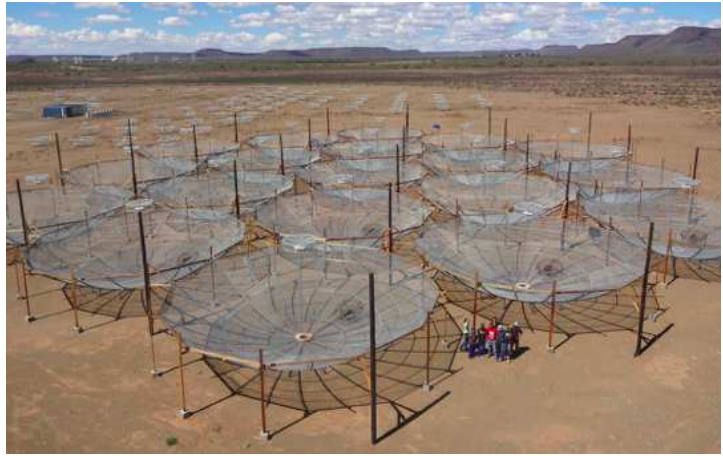}
\includegraphics[scale=0.5]{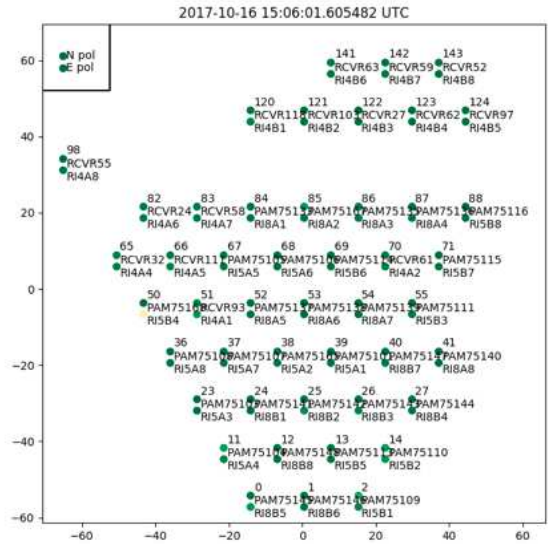}
%\vskip -0.9in
\caption{An aerial view of the first 19 antennas showing the grid pattern (Top) and positions and antenna numbers for HERA-47 in October 2017 (Bottom). The PAPER antennas and the MEERKAT array are visible in the background in the upper
picture. The lower figure is one of the standard diagnostic plots produced by the HERA on-line system.
Greendots indicate whether both polarizations were installed, and the antenna numbers are listed to the upper right of each antenna. The other number/letter combinations are internal diagnostic identifiers for electronic elements in the system. 
}
\label{fig:HERA47}
\end{figure}

Figure~\ref{fig:HERA47} shows the HERA layout for the data analyzed herein. This 
array had 47 antennas (henceforth we will refer to this configuration as 
HERA-47) populating grid locations as determined for the final HERA array. 
The final array will consist of 350 antennas in the full split-core 
grid, plus out-riggers to 1~km.
Our analysis employs commissioning data using prototype electronics. The system electronics will be upgraded in late 2018, including new broader-band feeds. The techinques developed herein
will be employed to characterize the new system, as it is deployed.

\subsection{Non-redundancy at low frequencies}\label{sec:nonred}

A key for HERA is the level of redundancy: how redundant are the 
redundant baselines in the grid array? 
Non-redundancy for a grid array could arise from cross-talk among cables, 
or in the correlator. 

Non-redundancy at low frequencies can also arise due to 
the fact that the sky brightness distribution
gets multiplied by the primary beam voltage 
pattern of each antenna. If these voltage patterns differ between antennas, 
then the product of voltage patterns will be different for each antenna pair, 
and hence the `effective sky' (product of the true sky and the antenna voltage 
patterns) will differ for each supposedly redundant baseline. If all the 
antennas were close to identical, then this should be a very minor effect.

In the classic antenna-based calibration process (for example, with the VLA or
ALMA), the signal is dominated by the calibrator sources in the center of the 
field. Hence, any differences in primary beam shape or sidelobe 
patterns is negligible, and the standard assumptions about antenna-based gain 
separability are valid.\footnote{In the old VLA broad band continuum correlator, 
closure errors arose due to differing bandpass shapes for the analog electronics 
for each antenna. This bandpass effect is directly analogous to what is being 
considered for the different primary beam power patterns herein. The new digital 
IF system for the JVLA has eradicated this problem.}

Antenna-based calibration for low frequency arrays is much more complex since 
the sky signal essentially fills the whole forward response of the telescopes 
(main beam and sidelobes); this signal consists of both the diffuse Galactic
emission and the extragalactic point sources. Hence, differences in the primary 
beam voltage patterns over the whole forward response factor directly into 
differences between measured visibilities for baselines that should be
redundant. Such voltage pattern differences could arise due to antenna
placement position errors, surface differences between antennas, blockage differences, 
or antenna feed positions and rotations.
Non-redundancy could also arise due to differing wide-field polarization response
between antennas \citep{smi11, bha01, bha03}. 

Likewise, if the field of view
is large, the ionosphere may show structure across the field of view. If these
structures are different for different antennas, this could
lead to non-closing effects. In the case of HERA-47, the field of view is about
$8^o$, which is larger than the expected sizes for significant structure in the ionosphere
of $\sim 1.5^o$ (corresponding to $\sim 10$km at the height of the troposphere 
\citep{kas07}). \replaced{However, the baselines considered herein are 29m or shorter, and hence
much shorter than the size of structures in the ionosphere, such that all the antennas
see the same ionosphere.}{
The short baselines considered here (29\,m and shorter), however, mean
that all the antennas will see the same ionosphere:  the diffracted antenna beams at the height of the atmosphere
(where the first Fresnel zone radius is around $\sqrt{\frac{\lambda h}{2\pi}}\sim 200\,$m) largely coherently  overlap, and 
the power  fluctuations of electron density at these short lenghtscales is small \citep{geh17}.
}
Hence, the summed effect of the ionosphere
should still cancel in a closure calculation. 

In cases of baseline-dependent responses for the interferometer, the standard calibration
assumption that the complex gains are factorize-able into an amplitude
and phase per antenna breaks down. This assumption is inherent to both
sky-based and redundant spacing calibration algorithms. In theory, the 
solution to this problem would be to measure the beams, then iterate through 
a `self-calibration to sky model' loop that includes direction dependent 
gains for all antennas \citep{bha08, smi15}.
In practice, such a process remains at the limit of computational abilities 
\citep{patil17}.

In summary, discrepancy of visibilities measured on geometrically redundant 
baselines for an array like HERA can arise due to numerous factors, including
boresight or global  voltage pattern differences, polarization response differences, 
and differences in the detailed shape of the voltage pattern away 
from boresight. At present, we have not determined which type of error is 
dominant for HERA. 

As a final note, if the non-redundancy is caused solely by antenna positions 
errors, then the closure calculation for each triad itself is still valid, 
but the assumed redundancy between triads breaks-down. 
One purpose of this paper is to investigate the closure phases for
redundant antenna triads in HERA, thereby getting an estimate of the
level of redundancy between supposedly redundant baselines.

\subsection{HERA-47 Data and Processing}\label{sec:data-processing}

We investigate closure phase spectra for HERA-47 data from October 2017
\citep[see also][]{car16a,car16b,car17}.
We analyze standard 10~min data sets in one linear polarization (the 'North'
polarization, as defined within the HERA project), for 
the following fields (Julian day and time). 

\begin{itemize} 

\item 2458043.12552: Observations when the Galactic Center was close to 
transit. The Galactic Center is at J2000 1745-2900. 

\item 2458043.24482: A field at RA = 21~hr with two bright sources. The 
two sources are: PMN J2107-2519 and PMN J2101-2802 with flux densities 
of 34~Jy and 22~Jy respectively at 150~MHz \citep{hur17}.

\item 2458043.58782: A field at RA = 05~hr with a bright cataloged source 
in the southern part of the beam. The source is PMN J0522-3628 with a flux 
density of  74.5~Jy at 150~MHz \citep{hur17}.

\end{itemize}
 
\noindent In all cases, the Sun was below the horizon.

The native Miriad format transit data from the correlator were converted 
to uvfits files using pyuvdata \citep{haz17}, 
including fringe-tracking on the 
zenith at the start of the 10~min data set. The record length for each visibility is 
10.7~s. The data were then imported into the Common Astronomy Software Applications
(CASA; \citep{mcm07}). Channel-based flagging 
was employed, based on visibility amplitude spectra.

We then performed self-calibration starting with a point source model
for the Galactic Center for delay calibration, and then bandpass
calibration using sky models derived from the data itself. This
process has been shown to converge 
to a sky brightness distribution consistent with previous low
frequency surveys, as has been documented in 
\citet{nik17}. We demonstrate below that the closure calculation is
independent of calibration, since the calibration process
makes strictly antenna-based corrections to the visibilities.
Note that direction dependent antenna-based gains may invalidate
the closure calculation, as discussed in \S\ref{sec:nonred}.
We do not analyze amplitudes herein, so the flux scale was
arbitrarily set to unity for the Galactic Center. We do perform a
rough flux scale bootstrap using the known sources below. 

We have written a new task in CASA called CCLOSURE\footnote{
http://www.mrao.cam.ac.uk/~bn204/soft/py/heracasa/}. This program generates 
{\it Python Numpy NPZ} format files with closure phases as a function of 
frequency for specified triads of antennas. We note that antenna ordering 
in the closure calculation is critical, i.e., for a given triad, going in 
opposite direction around the triad leads to a sign change. That is, 
$\phi_{123} = -\phi_{321}$. The CASA convention for a visibility phase is to 
always have the lower numbered antenna first. We have checked this sign 
convention, and the CCLOSURE task performs the proper re-ordering in the 
calculation, as set by the input parameters.  The output {\it npz} files 
contain information on the channel flags, the closure phases per channel per 
integration period, and the triads employed. We have written python 
scripts to read, plot, and analyze the closure spectra.

In the analysis below, we employ three types of triads: 

\begin{itemize}

\item EQ14 -- short equilateral triads made up of 14.6~m baselines. For 
example, (0,1,12). A total of 58 triads of this type are analyzed.

\item EQ29 -- longer equilateral triads of 29.2~m baselines. For example, 
(0,2,25). A total of 41 triads of this type are analyzed.

\item EWs -- short East-West triads. For example, (0,1,2). A total of 28 
triads of this type are analyzed.

\end{itemize} 

For the equilateral triads, there are two sub-types of redundant triads:
those pointing north such as (0,1,12), and those pointing south such as 
(11,12,1). It is easy to show that the closure phases of these sub-types
of north and south facing triads are identical (at least to the level 
of the redundancy of baselines), if calculated in opposite directions going 
around the triad (e.g., clockwise for north and counter-clockwise for 
south), while a calculation of the closure phase going in the same 
direction around the triads (eg. clockwise for both), leads to a sign change 
for north vs. south. The analogy for the linear triads would be analyzing, 
for example, (11,12,13) and (12,13,14).

When including all the redundant triads in the array of a given type as 
described above, many baselines may be included in more than one triad. Hence, 
the noise of the closure measurement averaged over redundant triads will not 
reduce strictly in an uncorrelated manner. However, there may be systematic 
effects that are reduced by including all the information, and in particular, 
the baselines on the border of the array that will not get included unless all 
triads are considered. 
 
\section{Results}

\subsection{Demonstration of calibration-independence of closure spectra}

\begin{figure}[htb]
\centering 
  \includegraphics[scale=1.1]{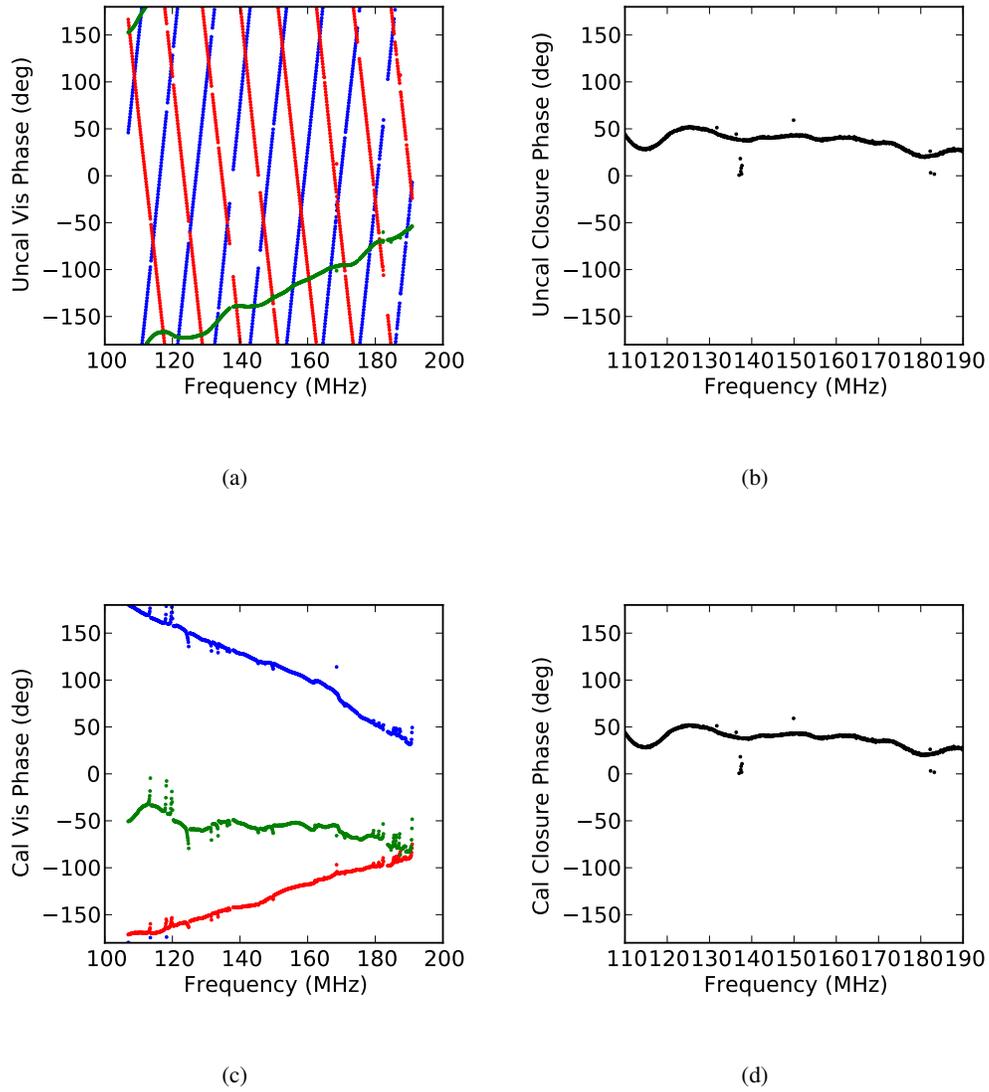}
%  \vskip -4.5in
  \caption{Left: Visibility phase spectra for three antennas in a short equilateral triad (Top: uncalibrated, Bottom: delay and bandpass calibrated). Right: Resulting closure phase spectrum. The measured closure phases are independent of antenna-based calibration of visibilities. These data correspond to the Galactic Center region.}
  \label{fig:vis}
\end{figure}

Figure~\ref{fig:vis} shows examples of visibility phases, and
closure phases, versus frequency for one EQ14 triad (121,122,141). Two sets 
are shown -- uncalibrated and calibrated data. 
The uncalibrated visibility phase spectra show large phase gradients,
indicative of large relative delays between antennas. These delays are
corrected in the  calibrated spectra.

The closure spectra for the uncalibrated and calibrated data are identical to 
within the numerical accuracy. While this result is as expected, given that 
sky-calibration is strictly an antenna-based correction, it remains reassuring 
that: (i) the processing itself does not somehow alter the answer, and (ii) we 
can employ either calibrated or uncalibrated data in the closure analysis and 
still obtain information on the true sky signal. In the analysis below, we 
employ the calibrated and flagged data sets, although, we emphasize that the 
results are unchanged if using uncalibrated data.

\subsection{Examples of closure spectra: three fields}

\begin{figure}[htb]
%\hspace*{-0.5in}
\centering
  \includegraphics[scale=0.8]{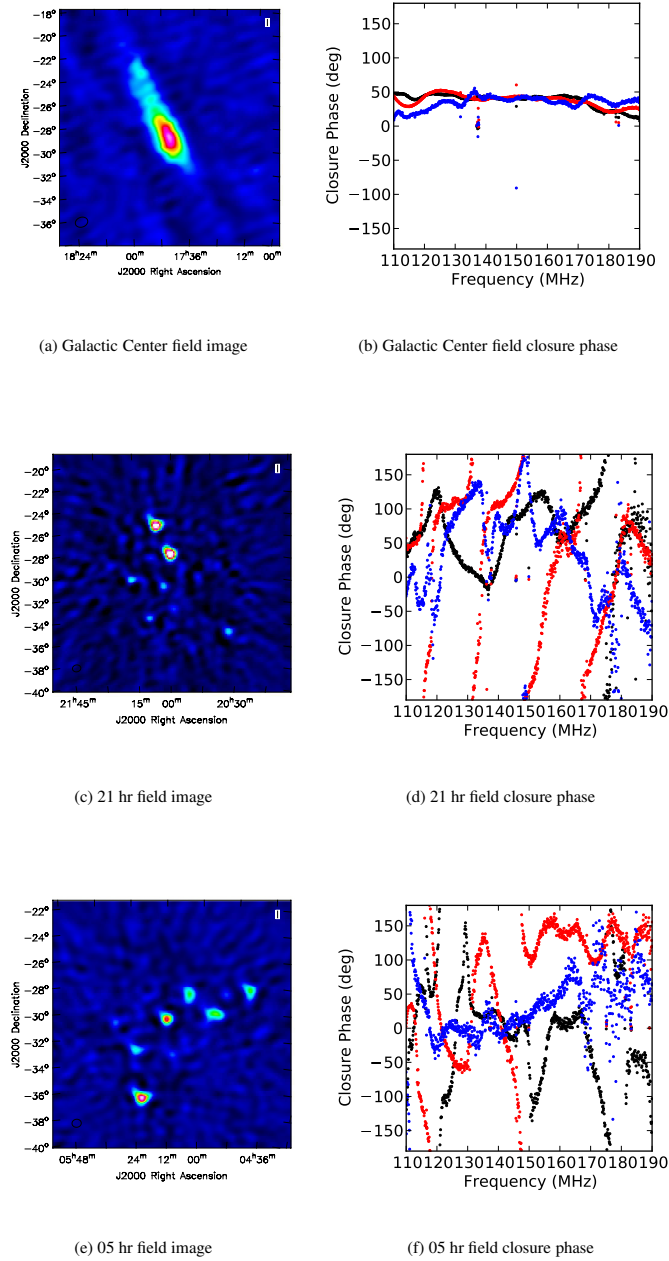}
%  \vskip -1.0in
  \caption{Top left: The HERA-47 image of the Galactic Center region at 150~MHz, with a synthesized beam of FWHM = $38'$. Top right: Closure phase spectra on three representative triads averaged over 1~min: black = short east-west (EWs), e.g., (0,1,2); red = 14m equilateral (EQ14), e.g., (0,1,12), and blue = 29m equilateral (EQ29), e.g., (0,2,25). Middle: Same, but for the 21~hr source field. Bottom: Same, but for the 05~hr field. Note that the next brightest four or five sources (yellow to red) in the 21~hr and 05~hr fields have plausible counterparts in the GMRT TGSS \citep{Int17}.}
  \label{fig:3triads}
\end{figure}

In Figure~\ref{fig:3triads}, we show images for the three fields analyzed in this
study, plus examples of the EWs, EW14, and EW29 closure spectra. The
sky calibration and imaging was performed as described in \citet{car16a,nik17}.  
Standard channel-based flagging was performed based on the visibility 
amplitude spectra.  

In brief, the calibration process starts with a delay and mean phase
offset calibration using a point source model for the Galactic Center.
A CLEAN model is then generated from the delay-calibrated Galactic
center data, and the complex bandpass (frequency-dependent, antenna-based 
gains), is then derived using this CLEAN model. The delay and bandpass
corrections are then applied to the visibility data for 
all three fields for final imaging. We imaged the data using CASA CLEAN, with a
Briggs weighting of the visibilities, with a robust parameter of 
-0.5 \citep{bri92}.
We have employed multi-frequency synthesis across the band 110--190~MHz. 
The images are not corrected for the power response of the primary beam.

The image of the Galactic Center shows the bright center of the
galaxy, and diffuse emission from the Galactic plane extending across
the primary beam. The 21~hr field shows clearly the two sources, PMN
J2107-2519 and PMN J2101-2802, each within a few arcmin of the
expected position (a small fraction of the synthesized beam FWHM of
38').  A few other sources are visible in this field. These are a
factor of few weaker than the two bright sources.  The 05~hr field
shows PMN J0522-3628 clearly. The next brightest four or five sources
(yellow to red) in the 21~hr and 05~hr fields have plausible
counterparts in the GMRT TGSS \citep{Int17}.

We have not performed a rigorous flux scale versus frequency
calibration for these data. However, we can do a rough scaling based
on the catalog flux density for the source PMN J2101-2801. This source
passes closest to the HERA zenith. This source is $2.6^\circ$ from
zenith in the image, or at the 79\% power point of the primary beam
(assuming an Airy disk for a 14~m antenna at 150MHz). Bootstrapping
from J2101-2801, implies an apparent source strength (uncorrected for
the primary beam) for PMN J2107-2526 of 16Jy.  This source is at the
34\% point of the primary beam, and the catalog flux density is 32Jy,
so the apparent flux should have been around 11Jy. For J0522-3628 the
relevant numbers are an apparent flux density of 13~Jy and an expected
apparent flux density (corrected by a factor of 0.22 due to the
primary beam) of 16~Jy. For the Galactic Center, the peak surface
brightness would then be 52~Jy~beam$^{-1}$, and the total in the field
is 350~Jy. We note the rms noise in the 05~hr field is
0.2~Jy~beam$^{-1}$.  Overall, the derived flux scaling is good to at
best 30\%, due to the lack of spectral corrections and detailed
knowledge of the primary beam. This scale is not relevant to the
closure phase analysis.

As for the closure spectra, it is clear that the frequency structure
becomes more pronounced with longer baselines, and more complex
sources. For instance, the Galactic Center field is dominated by the
main beam emission composed of the bright center of the galaxy, and 
the smooth extended Galactic plane attenuated by the
primary beam. This field shows the smoothest closure spectra.
Conversely, the PMN J0522-3628 field has at least six sources of
comparable flux density (5 to 10~Jy), and has more pronounced
structure in the closure phase spectra. Note these point sources
comprise at most 50\% of the total flux density in the shorter
spacings of the array. 

\subsection{Redundancy}

One of the main goals of this study is to provide a metric for non-redundancy of the 
array.  We have considered both a mean and median analysis when summing 
closure spectra for redundant triads, and find that the results are the same to within the noise,
while the median absolute deviation is comparable to the rms deviation. Given that
the median statistic is generally more robust to outliers, we conservatively
adopt the median statistic in the analysis below.

\begin{figure}[htb]
%\hspace*{-1.0in}
\centering
  \includegraphics[scale=0.85]{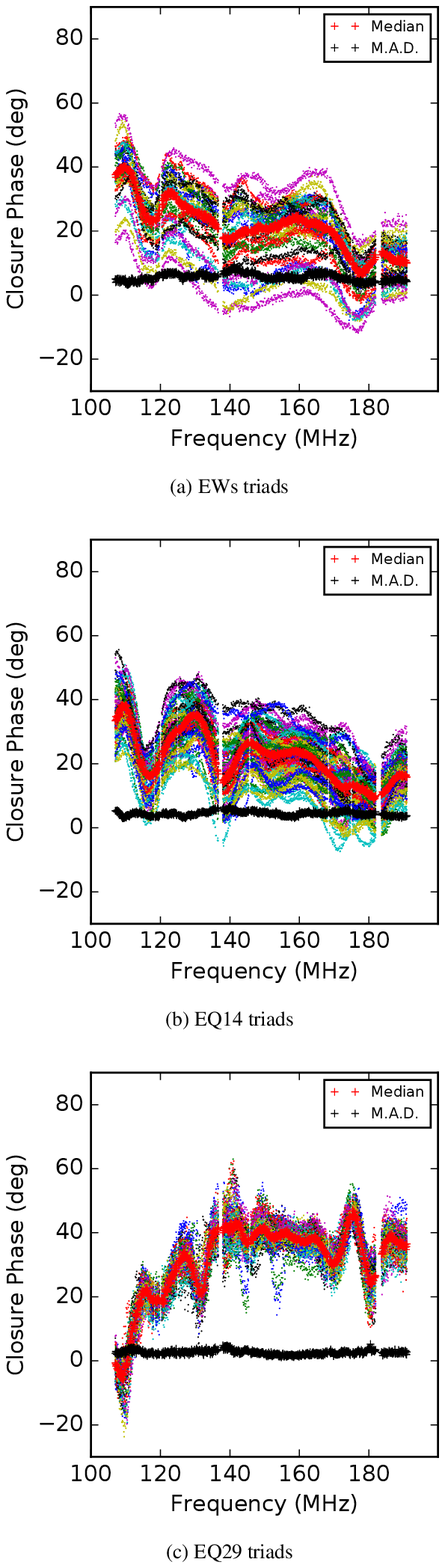}
%  \vskip -1.1in
  \caption{Top: Median (bold red) and median absolute deviation (black) for closure phase spectra on the GC field for EWs triads, calculated across all redundant triads of this type, and for spectra averaged over 1~min. The data for each triad is also shown independently in lighter colors, to demonstrate the scatter between redundant triads. Middle: Same, but for EQ14 triads. Bottom: Same, but for EQ29 triads.}
  \label{fig:medmad}
\end{figure}

Figure~\ref{fig:medmad} shows the full distribution of closure spectra for all
redundant triads of types EQ14, EQ29, and EWs, averaging over 10 records for GC data. 
We use the GC data to ensure high signal-to-noise over the full spectral
range. We also include both up and downward triads for the EQ, calculated in such a 
way as to sum rather than difference (see \S\ref{sec:data-processing}), as well 
as all the EW short triads. Including all triads ensures that all antennas on
the boundaries of the array are included in the summation. 

Figure~\ref{fig:medmad} also shows the median for all triads, and the median 
absolute deviation (approximately the rms scatter at a given frequency), as a
function of frequency. While the qualitative behaviour for the redundant
triads is similar, it is clear that the systematic difference between triads
is much greater than the noise within a given triad. The critical point is 
that, if all our baselines were exactly redundant, then the curves for 
redundant triads in Figure~\ref{fig:medmad} would be coincident.  Any 
non-coincidence indicates departures from redundancy.  

\begin{figure}
\centering 
\includegraphics[width=\textwidth]{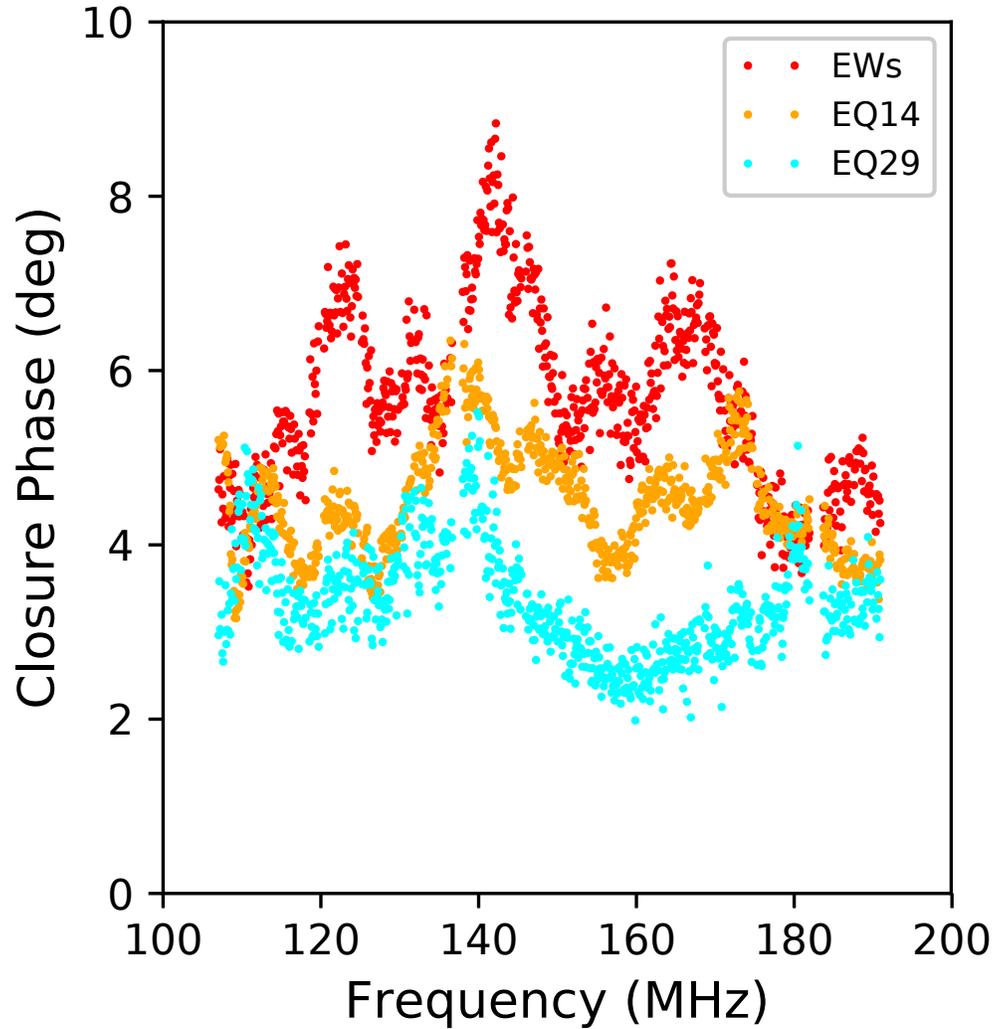}
\caption{The median absolute deviation between redundant triads versus frequency from figure 7. The three triads
are show in different colors. The deviations for the EQ29 triads are smaller than the EQ14 triads. 
Larger deviations from redundancy for longer baseline triads may be due to the longer baselines being more sensitive to bright sources in the main beam, where antenna to antenna beam differences may be less than for the more distant sidelobes of the antenna beam. The shorter triads have a larger contribution from diffuse emission, which fills both the main beam and the far sidelobes, although this phenomenon is likely field-dependent.
}
\label{fig:mad}
\end{figure}

Figure~\ref{fig:mad} shows a magnified view of the median absolute deviation of the 
closure spectra for the three triads in Figure~\ref{fig:medmad}. 
The median absolute deviation is largest for the EWs triad (mean over
the frequency range of about $5.8^o$), dropping to about $4.5^\circ$ for EQ14, 
and $3.5^\circ$ for EQ29. There are local peaks by factors
of two, but the mean value does not systematically change with frequency.
It is interesting that, even though the intrinsic structure in the closure
spectra is more complex for the longer EQ29 triads, the departures from redundancy 
are smaller than for EQ14.  This decrease of deviation with longer baselines
may indicate an increasing dominance of the brighter sources in the main
beam to the longer baseline visibilities, 
although such differences clearly depend on a specific field 
brightness distribution, ie. the location and prominence of strong point
sources relative to the diffuse continuum. Mutual coupling may also be 
less for longer baselines.

A point to keep in mind is that any standard antenna-based calibration process, such as 
redundant calibration or hybrid mapping sky calibration, will not correct 
for these differences in the closure spectra, i.e., these departures from 
redundancy will lead to `closure errors' in the antenna-based calibration 
process \citep{per99}.

We take $4.5^\circ$ as a representative rms departure from closure redundancy 
between triads in HERA. We consider how such an error might propagate into 
visibility errors using antenna-based calibration techniques.  We make the 
simple assumption that the rms phase variations of the visibilities due to 
e.g., differing primary beams ($\phi_{rms,vis}$), would add in quadrature 
when calculating closure phase.  In this case, the rms phase variations for 
the closure phases ($\phi_{rms,1,2,3}$), would be: 
$\phi_{rms,1,2,3}^2 = \phi_{rms,1,2}^2 + \phi_{rms,2,3}^2 + \phi_{rms,3,1}^2$. 
A value of $\phi_{rms,1,2,3} = 4.5^\circ$ then translates into: 
$\phi_{rms,vis} = \phi_{rms,triad}/(3^{1/2}) = 2.6^\circ$.

How will non-closing calibration errors of this magnitude affect imaging?  
The standard relation for image dynamic range due  calibration phase errors 
is given in \citet{per99}:
$DR \sim N/\phi_{rms,vis}$, where $N$ is the number of antennas and 
$\phi_{rms}$ is in radians. For HERA350, the implied image dynamic range 
limit due to non-closing errors in antenna based calibration schemes is 
$DR \sim 8000$. 
For comparison, the highest dynamic range imaging
using eg. LOFAR and the VLA has reached a level of order 10$^5$.
However, such studies have been limited to specific fields where the emission 
is dominated by a strong point source, and in which self-calibration can be
performed to the level of hundredths of a degree\citep{ber10, per17}.
We re-emphasize that our analysis employs commissioning data using prototype electronics. The electronics of the array, including
the feeds, will be 
upgraded in late 2018, after which a reanalysis will be performed.

\subsection{Screening ill-behaved antennas}

Closure calculations can also be used to find ill-behaved antennas, such as 
antennas in which the cables for each polarization where inadvertently swapped, 
in the absence of calibration. Admittedly, such 
a cross-polarized antenna would also show very low amplitudes, but it
remains possible the low amplitudes are due to a low gain somewhere in
post-front end signal chain, and hence might normalize-out after 
calibration. Closure redundancy departures are independent of calibration.

\begin{figure}[h]
\centering 
\includegraphics[width=\textwidth]{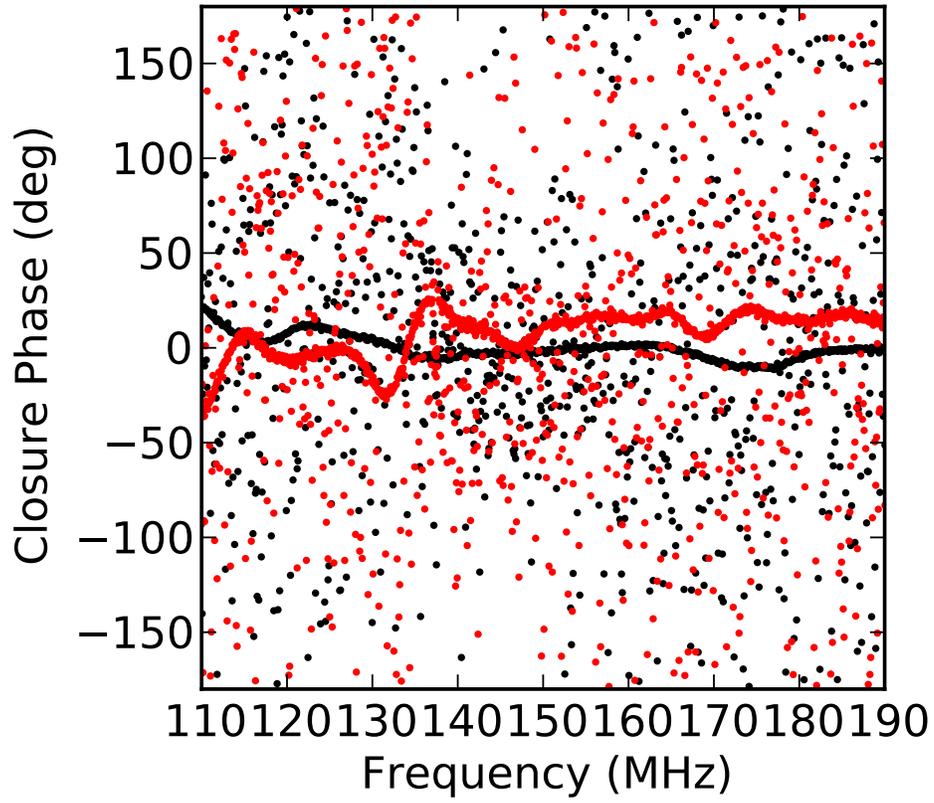}
\caption{The black curve shows the EWs linear triad closure spectrum for the triad (51,52,53), while the black points show a similar curve for EWs triad (50,51,52). The red curve is for a short linear triad (51,67,84), but oriented 30$^\circ$ from north, and the red points are for a linear triad of the same orientation but formed from antennas (50,66,83). These are for Galactic Center data with 60~s averaging. Antenna 50 was found to be cross-polarized and hence triads that include this antenna show noise-like closure phase spanning $\pm 180^\circ$.
}
\label{fig:badants}
\end{figure}

Figure~\ref{fig:badants} shows the two sets of short linear triads, both EW 
and oriented 30$^\circ$ from north. The linear east-west triads are: (50, 51, 52) 
and (51, 52, 53) in black and gray respectively, and the linear triads oriented 
at 30$^\circ$ are (50, 66, 83) and (51, 67, 84) in orange and red 
respectively. Triads containing antenna 50 show much higher noise than those 
that do not. This behaviour was seen in the initial analysis of the data, 
and it was later found that antenna 50 is cross-polarized. The 
cross-polarized triads are essentially show noise-like closure phase spanning $\pm 180^\circ$.   

The main result is that cross-polarized, or otherwise ill-behaved 
antennas, are very obvious in an analysis of redundant closure spectra,
independent of calibration.

\subsection{UTC Variation}\label{sec:UTC-variation}

We next investigate the time variation of the closure spectra over
a 10~min observation. The time variation is important in setting the 
coherent averaging time on which we can perform our cosmological analysis 
(Thyagarajan et al. 2018). Time variation is due to the sky moving through
the primary beam of the transit instrument. We note that the sky moves
$\simeq 13^\circ$ at zenith in 1~hour at HERA latitude, and that the FWHM of 
the HERA beam is $8^\circ$ at 150~MHz. 

\begin{figure}[htb]
%  \hspace*{-1.0in}
  \centering 
  \includegraphics[scale=0.9]{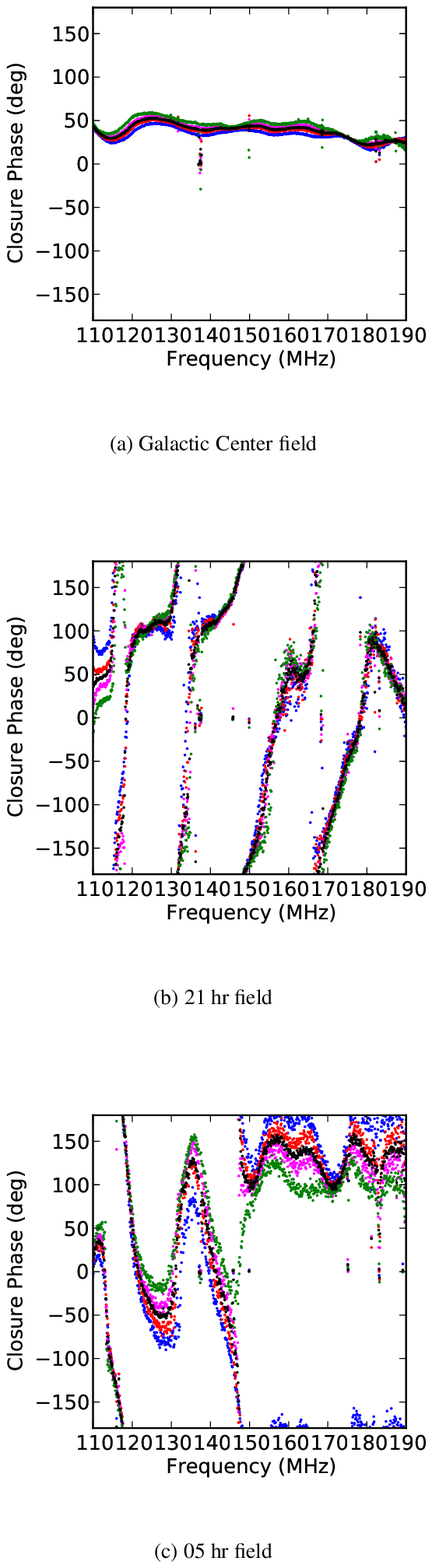}
%  \vskip -1.4in
  \caption{Top: Closure spectra for an EQ14 triad on Galactic Center data. The red-blue-magenta-green curves are consecutive 2.5~min averaged spectra. The black is the spectrum averaged over 10~min. Middle: Same, but for the 21~hr field. Bottom: Same, but for the 05~hr field. The variation across 2.5~min intervals is found to exceed intrinsic scatter in each of the intervals.}
  \label{fig:UTC-variation}
\end{figure}

Figure~\ref{fig:UTC-variation} shows closure spectra for EQ14, EQ29,
and EWs for four consecutive times averaged over 2.5~min each, plus 
the 10~min averaged closure spectra. There is clear variation of the closure
spectra from one 2.5~min time to the next, much larger than the intrinsic
scatter. 

To investigate this time variation more quantitatively, we have
performed an Allan Variance analysis of EQ14 triads on the three
different fields (GC, 21~hr, 05~hr).  Allan Variance is a
standard way of analysing stability of measurement when the
instabilities or noise affecting the measurement are non-Gaussian
distributed.  An introduction to Allan Variance is given by
\cite{tho17} and an example its application to quantifying stability
of astronomical observations is \cite{2001A&A...373..746S}.  We
computed the Allan deviation of the observed closure phase spectra in
order to estimate the time period over which they can be usefully
averaged, i.e., before the rotation of the sky introduces a drift
larger than the thermal-like noise in the averaged spectrum.  The
deviation was computed using the overlapping method (see equation 9
and 11, NIST SP 1065 "Handbook of frequency stability analysis") with
integration time equal to the stride period.

\begin{figure}[htb]
%  \hspace*{-0.3in}
\centering
  \includegraphics[scale=0.8]{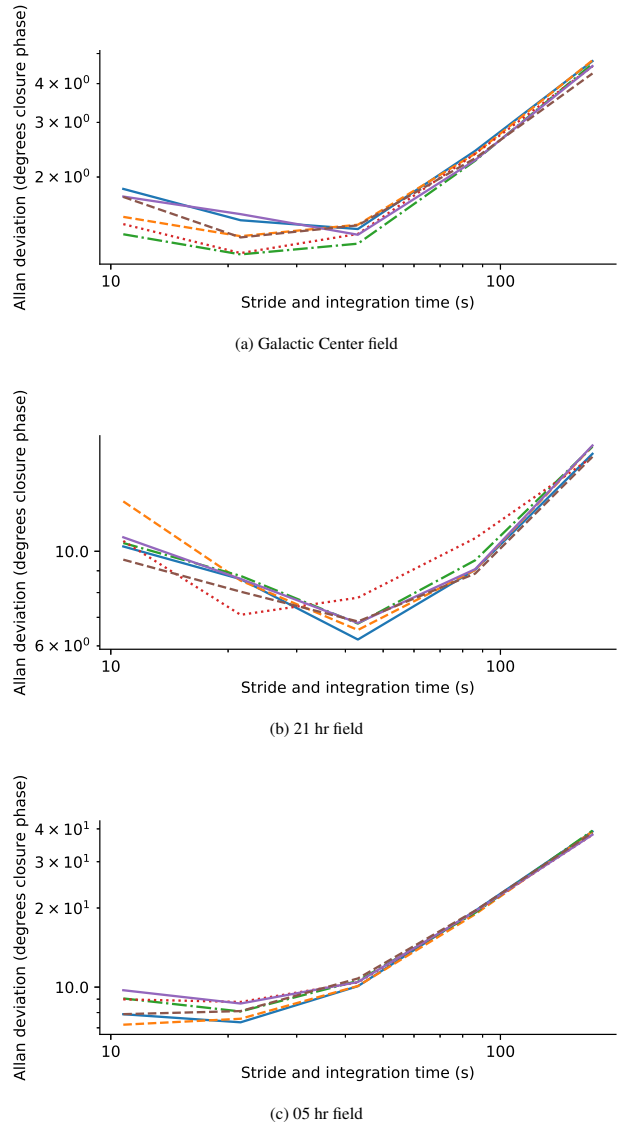}
  \vskip -1.1in
  \caption{Allan standard deviations of the closure phase on the East-West triad for: (top panel) the Galactic Centre field; (middle panel) the 21h field and (lower panel) the 5h field. Each panel shows the Allan standard deviation for 6 neighbouring frequency channels (shown by the 6 lines in each panel) at around 140\,MHz. Beyond a timescale of 20--80~s, the difference caused by the drifting of sky becomes larger than the noise level in closure phases, which are dominant on shorter timescales.}
  \label{fig:ASD}
\end{figure}

The results are shown in Figure~\ref{fig:ASD} for the smallest equilateral
triad for three different fields. We find a shallow minimum in all cases
for delays around 20~sec to 80~sec, beyond which the  Allan deviation increases
steadily. The correlator dump time of HERA is 10.7s. 

\subsection{LST Variation}

The time variations seen in \S\ref{sec:UTC-variation} are due to the sky moving
through the beam of the transit array. We now investigate variations
from one day to the next, but at the same LST, such that the sky
position is identical.

\begin{figure}[htb]
%  \hspace*{-0.5in}
  \centering 
  \includegraphics[scale=0.9]{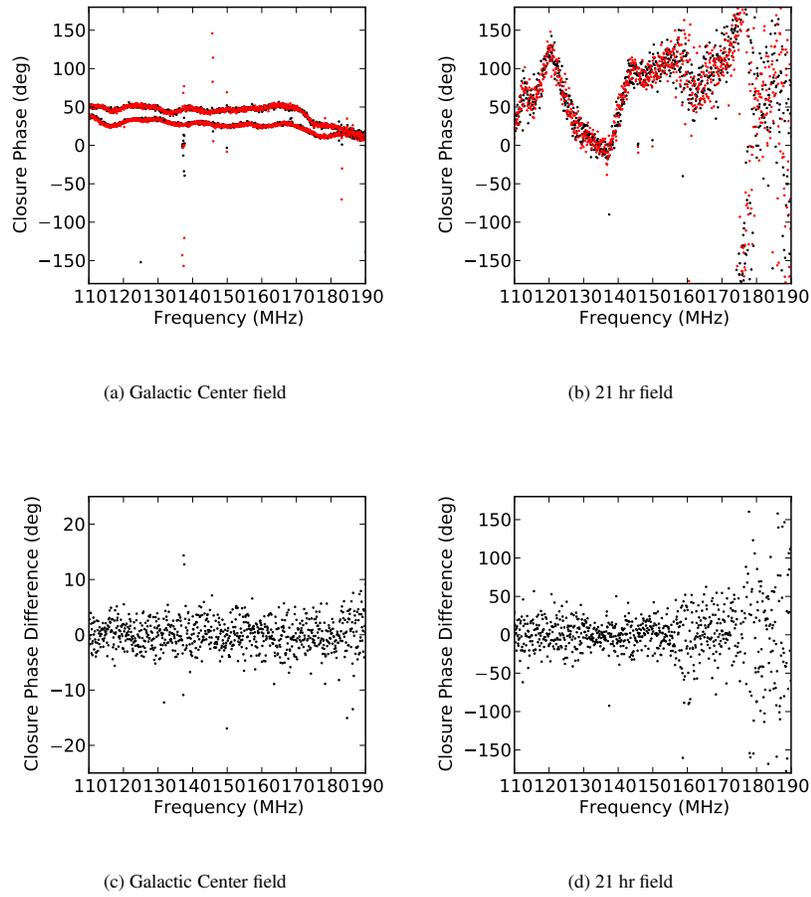}
%  \vskip -3in
  \caption{Top left: Closure spectra for an EWs triad for GC data for a 10.7~s record, from the same LST on two consecutive days. Two examples are shown, separated by 8~min. The red and black curves are for day 1 and 2. Top right: same but for the 21~hr field. Bottom left: Difference between closure phase spectra from two 10.7~s records on consecutive days for the GC data and an EWs triad. Bottom right: Same, but for the 21~hr field. The difference between closure phases at same LST on consecutive days is consistent with expected closure phase noise rms.}
  \label{fig:LST-variation}
\end{figure}

Figure~\ref{fig:LST-variation} shows the closure phase spectra for a single 
10.7~s record at the same LST (within 0.4~s), on two consecutive days for the 
GC observations and the 21~hr source observations. These spectra are for an 
EWs triad. The red and black curves are for the two different days. 
The curves overlap to well 
within the noise in each case. We then difference the curves from day 1 and 
day 2. The difference plots are consistent with zero, within the noise. 
For the GC  data we show two examples, separated by 8~min. 

Hence, the closure phase spectra appear to be stable at a given LST from one
day to the next, at least for the limited amount of data we have analyzed.  
Stability is critical for coherent averaging of
the closure phase spectra from multiple days.

\subsection{Closure Amplitudes}\label{sec:clamp}

For completeness, we calculate closure amplitudes. Closure amplitudes
($CA$),  involve the product of four visibility amplitudes ($A_{ij}$): 

$ CA_{ijkl} = [A_{ij} A_{kl}]/[A_{ik} A_{jl}] $

\noindent As with closure phase, the antenna based calibration terms in this
product of visibilities cancel, and one is left with a `true sky
measurement', even for uncalibrated visibilities 
\citep[see equation 10.44 in][]{tho17}.
The difficulty with analyzing closure amplitudes is that the calculation 
involves a division, which can lead to divergences around nulls in the 
visibility functions, for noisy data.

\begin{figure}[htb]
  \centering
  \subfloat[][Closure amplitudes]{\label{fig:GC-clamp}\includegraphics[width=0.45\linewidth]{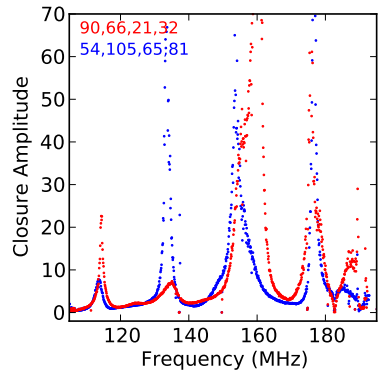}} \\
  \subfloat[][Visibility amplitudes]{\label{fig:GC-visamp}\includegraphics[width=0.45\linewidth]{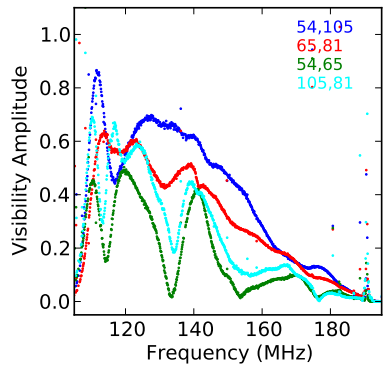}} \\
  \caption{Top: Closure Amplitude spectra for two short, redundant quadrangles for GC data from the HERA19 configuration \citep[see][]{car16a}.  We have truncated the maximum closure amplitude in the figure.
The closure amplitude magnitudes go up to 400.
Bottom: The calibrated visibility amplitude spectra for one set of four antennas used in the closure amplitude spectrum. The frequency range is 100MHz to 200MHz. The high-valued closure amplitudes in the spectra are caused by and coincide with low-valued visibility amplitudes.}
  \label{fig:clamp}
\end{figure}

Figure~\ref{fig:clamp} shows the closure amplitudes for two short, redundant 
quadrangles. These are data from the older HERA-19 array \citep{car17,car16a}.. 
There is broad similarity between the two quadrangles, but there are large 
peaks in the distributions that do not replicate well between quadrangles. 
We have truncated the maximum closure amplitude in the figure.
The magnitude of the closure amplitudes go as high as 400.

Figure~\ref{fig:clamp} also shows the visibility amplitude spectra for the 
baselines in one of these quadrangles. The peaks in the closure amplitude
spectra occur when one or more of the visibility spectra approach zero. 
These figures demonstrate clearly that the closure amplitudes diverge if 
one of the visibility amplitudes approaches zero.

Two methods may facilitate use of closure amplitudes in the search for
the reionization signal. The simplest would
be to avoid spectral regions in which one or more visibility spectra approach
a null. A related method, in the absence of a wide enough spectral range free
of nulls, is to perform an amplitude weighted power spectral analysis, thereby
down-weighting spectral regions with low visibility amplitudes.

\section{Discussion}

We investigate closure phase spectra in the context of the massively
redundant, low frequency transit array, HERA.  The primary purpose of this 
work is as supporting material for the potential application of closure 
phase spectra as a method to detect the \HI\ 21~cm signal from the 
neutral intergalactic medium during Cosmic Reionization. We consider two 
important parameters in this regard: temporal stability and array 
redundancy.  More generally, HERA is an excellent laboratory to study the 
nature of closure and redundancy in interferometry.

We show that, as expected, the closure phase measurements are conserved 
before and after antenna based calibration, to within the numerical noise. 
For the fields examined, the closure spectra are relatively smooth with time 
and frequency (structure on scales of tens of MHz).  Spectral complexity 
increases with complexity of structure in the field and baseline length.

We find that the median absolute deviation of the closure phase
measurements between redundant triads in HERA-47 data is about
4.5$^\circ$, and that this deviation is dominated by triad-to-triad
variation, and not signal-to-noise.  The deviations from redundancy
increase from the long EQ to the short EQ triads, even though the long
baseline triads show more intrinsic structure in the closure
spectra. This decrease of deviation with longer baselines may indicate
an increasing dominance of the brighter sources in the main beam to
the longer baseline visibilities, although this could be field
dependent, or due to decreased cross coupling for longer baselines.

The measured deviations provide  
a rough estimate of the level of redundancy between redundant baselines in 
HERA. On average, the rms phase deviation between redundant visibilities 
is about $2.6^\circ$.  These errors fold directly into calibration errors 
using standard antenna-based calibration algorithms.  We estimate errors of 
this magnitude would limit the imaging dynamic range of HERA-350 to 
$\sim 8000$ using standard antenna-based calibration techniques. 
We show that closure phase spectra of redundant triads can be used to identify 
bad antennas, independent of calibration. In our case study, the identified 
antenna was found to be cross polarized.  

Departures from redundancy 
will clearly affect antenna-based calibration schemes (ie. redundant calibration
and sky-based calibration), and affect the
power spectral analysis when averaging either redundant visibility spectra or closure spectra. 
Identifying and removing the worst antennas will help in this process. We emphasize
that the analysis presented herein is for a small subset of the HERA antenna,
using commissioning electronics for the system. Antennas construction is proceeding to
the full 350 complement of antennas,
and in late 2018 the entire feed, amplification, data transmission, and back-end
system of HERA will be upgraded.  Future studies will focus on determining the
level of redundancy, using closure techniques, among others, and how departures
from redundancy affects the calibration and final cosmological measurement.

We consider the time variation of the closure spectra over the course
of an observation. This variation is dominated by the sky moving
through beam of the transit telescope. An Allan variance analysis
requires an averaging time of less than two minutes, to avoid
systematic changes due to sky transit exceeding visibility noise.  We
also consider changes from day to day at a fixed LST. In this case,
the variations are extremely small, much less than the noise in the
spectra, and LST binning from day to day will be effective,
due to the stable system. 

We investigate closure amplitudes. While the behavior for redundant
quadrangles are similar, the calculation of closure amplitude involves 
division, and the values can diverge in regions when one or more
of the visibility spectra approaches zero. 

In the following papers in the series, we will employ the analysis
tools derived herein to the question of \HI\ 21~cm cosmology\citep{tcn18}. In
particular, we will attempt to use closure spectra as a robust means
to detect the \HI\ 21~cm emission from cosmic reionization,
independent of antenna-based calibration and calibration errors, and
the ionosphere. 

\acknowledgments
The National Radio Astronomy Observatory is a facilities of the
National Science Foundation operated under cooperative agreement by
Associated Universities, Inc. Carilli \& Bernardi acknowledge support
from a Newton Fund grant from the UK Royal Society.
This work was supported by the U.S. National Science
Foundation (NSF) through awards AST-1440343 and AST-1410719.
The closure phase data used in the plots can be found at: 
\verb|http:www.aoc.nrao.edu/~ccarilli/ccarilli.shtml|

%% ------------------------------------------------------------------------ %%
%% Citations

% Please use ONLY \citet and \citep for reference citations.
% DO NOT use other cite commands (e.g., \cite, \citeyear, \nocite, \citealp, etc.).

%% Example \citet and \citep:
%  ...as shown by \citet{Boug10}, \citet{Buiz07}, \citet{Fra10},
%  \citet{Ghel00}, and \citet{Leit74}. 

%  ...as shown by \citep{Boug10}, \citep{Buiz07}, \citep{Fra10},
%  \citep{Ghel00, Leit74}. 

%  ...has been shown \citep [e.g.,][]{Boug10,Buiz07,Fra10}.

%%  REFERENCE LIST AND TEXT CITATIONS
%
% Either type in your references using
%
%\bibliography{refs}

\clearpage
\newpage

\clearpage
\newpage

\clearpage
\newpage

\clearpage
\newpage

\clearpage
\newpage

\clearpage
\newpage

\clearpage
\newpage

\clearpage
\newpage

\clearpage
\newpage

\clearpage
\newpage

\clearpage
\newpage

\clearpage
\newpage

\end{document}